\documentclass[prl,aps,superscriptaddress,footinbib,twocolumn]{revtex4-2}

\usepackage{graphicx}
\usepackage{dcolumn}
\usepackage{bm}
\usepackage[unicode=true,bookmarks=true,bookmarksnumbered=false,bookmarksopen=false,breaklinks=false,pdfborder={0 0 1},backref=false,colorlinks=true]
{hyperref}
\hypersetup{
	linkcolor=magenta, urlcolor=blue, citecolor=blue, pdfstartview={FitH}, hyperfootnotes=false, unicode=true}
\usepackage{url}
\usepackage{xcolor}

\usepackage[utf8]{inputenc}
\usepackage{amsmath}
\usepackage{amssymb}
\usepackage{graphicx}
\usepackage{esint}
\usepackage{ulem}
\usepackage{multirow}

\begin{document}

\title{Coherent control of quantum topological states of light in Fock-state lattices}

\affiliation{Interdisciplinary Center for Quantum Information,~State Key Laboratory of Modern Optical Instrumentation,~and Zhejiang Province Key Laboratory of Quantum Technology and Device,~School of Physics, Zhejiang University, Hangzhou 310027, China\\}
\affiliation{Hangzhou Global Scientific and Technological Innovation Center, Zhejiang University, Hangzhou 311215, China\\}
\affiliation{CAS Center of Excellence in Topological Quantum Computation, Beijing
100190, China}

\author{Jinfeng Deng$^{1,*}$, Hang Dong$^{1,*}$, Chuanyu Zhang$^1$, Yaozu Wu$^1$, Jiale Yuan$^1$, Xuhao Zhu$^1$, Feitong~Jin$^1$, Hekang~Li$^1$, Zhen Wang$^1$, Han Cai$^1$, Chao Song$^{1,\dagger}$, H. Wang$^{1,2,\S}$, J.Q. You$^1$, and Da-Wei Wang$^{1,3,\ddagger}$}

\date{\today}

\begin{abstract} 

Topological photonics provides a novel platform to explore topological physics beyond traditional electronic materials and stimulates promising applications in topologically protected light transport and lasers.
Classical degrees of freedom such as polarizations and wavevectors are routinely used to synthesize topological light modes. Beyond the classical regime, inherent quantum nature of light gives birth to a wealth of fundamentally distinct topological states, which offer topological protection in quantum information processing. Here we implement such  experiments on topological states of quantized light in a superconducting circuit, on which three resonators are tunably coupled to a gmon qubit. We construct one and two-dimensional Fock-state lattices where topological transport of zero-energy states, strain induced pseudo-Landau levels, valley Hall effect and Haldane chiral edge currents are demonstrated. Our study extends the topological states of light to the quantum regime, bridges topological phases of condensed matter physics with circuit quantum electrodynamics, and offers a new freedom in controlling the quantum states of multiple resonators.
\end{abstract}

\maketitle

The quantum Hall effect \cite{Klitzing1980} reveals new phases of matter classified by the topological invariants of energy bands \cite{Thouless1982}. For two-dimensional electrons in strong magnetic fields, the chiral edge states between Landau levels contribute to the quantized Hall conductivity, which is immune to local defects. This topological effect can also exist without Landau levels, such as in the Haldane model \cite{Haldane1988}, which lays the basis for topological insulators \cite{Kane2005}. The optical simulation of quantum Hall edge states \cite{Haldane2008} opens a new research area, topological photonics \cite{Lu2014,Khanikaev2017,Ozawa2019}, which brings a wealth of applications in routing and generating electromagnetic waves, such as backscattering-free waveguides \cite{Wang2009d} and topological insulator lasers \cite{Bandres2018}. While quantum wave and fermionic statistics play a fundamental role in topological phases of electrons, the topological modes in photonic lattices are purely classical \cite{Ozawa2019}. Intriguingly, novel topological states emerging from light quantization and bosonic statistics have been predicted beyond classical interpretation \cite{Cai2020}. Recent development in circuit quantum electrodynamics (QED) \cite{Carusotto2020} makes it possible to realize these intrinsic quantum topological states of light, which provide quantum degrees of freedom in engineering photonic topology \cite{Cai2020,Yuan2021} and offer topological control knobs in bosonic quantum information processing \cite{Wang2016,Gao2019,Hu2019,Ma2021}.

\begin{figure}[t]
	\includegraphics[width=3.4in]{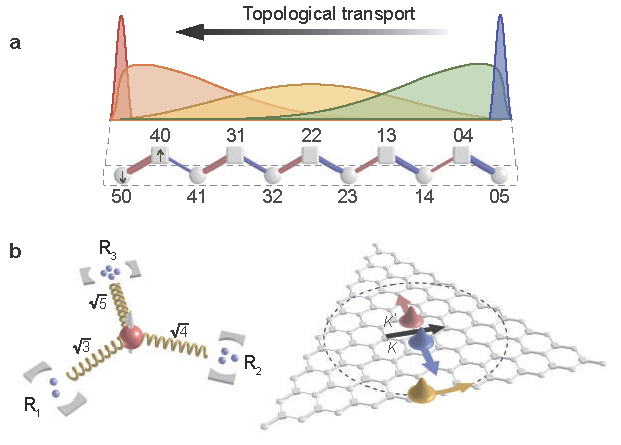}
	\caption{\label {Fig1} {\bf Fock-state lattices of multimode Jaynes-Cummings models.} (a) Topological transport of the zero-energy state of the SSH FSL with $N=5$. 
    The sublattice sites with the spin states $s = \uparrow$ ($\downarrow$) are denoted by squares (circles).
    Each site is labelled by $n_1n_2$, the photon numbers in R$_1$ and R$_2$. The thicknesses of the lines connecting neighboring sites are proportional to the coupling strengths $t_1$ (red) and $t_2$ (blue).
	The wave functions of four zero-energy states which inhabit the $\left|\downarrow\right\rangle$ sublattice during the topological transport are schematically drawn with different colors. (b) The schematic valley Hall response and the Haldane chiral edge state in a 2D FSL with $N=10$. An excited qubit is coupled to three resonators with different photon numbers $n_j$ (left). The coupling strengths are proportional to $\sqrt{n_j+1}$, which introduces competition between resonators to exchange photons with the qubit. All such Fock states with the same $N$ are coupled by the JC Hamiltonian to form a honeycomb lattice (right). The inhomogeneous coupling strengths induce an effective magnetic field in the 2D FSL. The VHE is featured by the wave functions at the two valleys moving in opposite directions perpendicular to an applied force (the black arrow). A Lifshitz topological edge (dashed line) separating semimetal and insulator phases locates on the incircle, which can host the Haldane chiral edge states (yellow wave function with the arrow showing the moving direction).}
\end{figure}

Instead of a lattice of resonators or waveguides in topological photonics, the topological states of quantized light are embedded in lattices of Fock states $\Pi_i |n_i\rangle$ with $n_i$ being the photon number in the $i$th mode. In the Fock-state lattice (FSL) a light mode provides a dimension \cite{Wang2016,Cai2020,Yuan2021}, in contrast to a lattice site in photonic lattices. Therefore, FSLs have the advantage of building high dimensional lattices with only a few cavity modes. To sketch such dimensional scalability we use the famous Jaynes-Cummings (JC) model \cite{Jaynes1963}, which describes the interaction between a two-level atom with quantized light. However, here we use multiple quantized light modes to couple the atom. With two light modes the Fock states form one-dimensional (1D) lattices of the Su-Schrieffer-Heeger (SSH) model (see Fig.~\ref{Fig1} (a)) \cite{Su1979}. By just adding another mode, we obtain two-dimensional (2D) strained honeycomb lattices (see Fig.~\ref{Fig1} (b)) \cite{Guinea2010}. These lattices are featured by site-dependent coupling strengths, which originate from the property of the bosonic annihilation operator $a$,
\begin{equation}
a|n\rangle=\sqrt{n}|n-1\rangle.
\label{a}
\end{equation}
For the vacuum state, $a|0\rangle=0$, which leads to natural edges of FSLs when the photon number in one of the cavities reduces to zero. More interestingly, {FSLs also have topological edges which host zero-energy states, resulted from the competition between resonators in exchanging photons with the atom.} Such a simple mechanism enables FSLs to realize several most important models in topological physics, in particular the seminal SSH and Haldane models, which have been the focus in various quantum platforms \cite{Meier2016,Kiczynski2022,Groning2018,Sylvain2019,Jotzu2014,Roushan2014,Zhao2022}.~Here we demonstrate adiabatic transport of topological zero-energy states in 1D SSH FSLs, where Fock states are topologically transferred from one cavity to another while maintaining the quantumness in superposition states. In 2D FSLs we observe the valley Hall effect (VHE) \cite{Xiao2007} and the Haldane chiral edge current \cite{Rechtsman2013}, which offer a topological route of engineering quantum states of multiple resonators. 

\begin{figure*}[!htp]
		\includegraphics[width=7.0in]{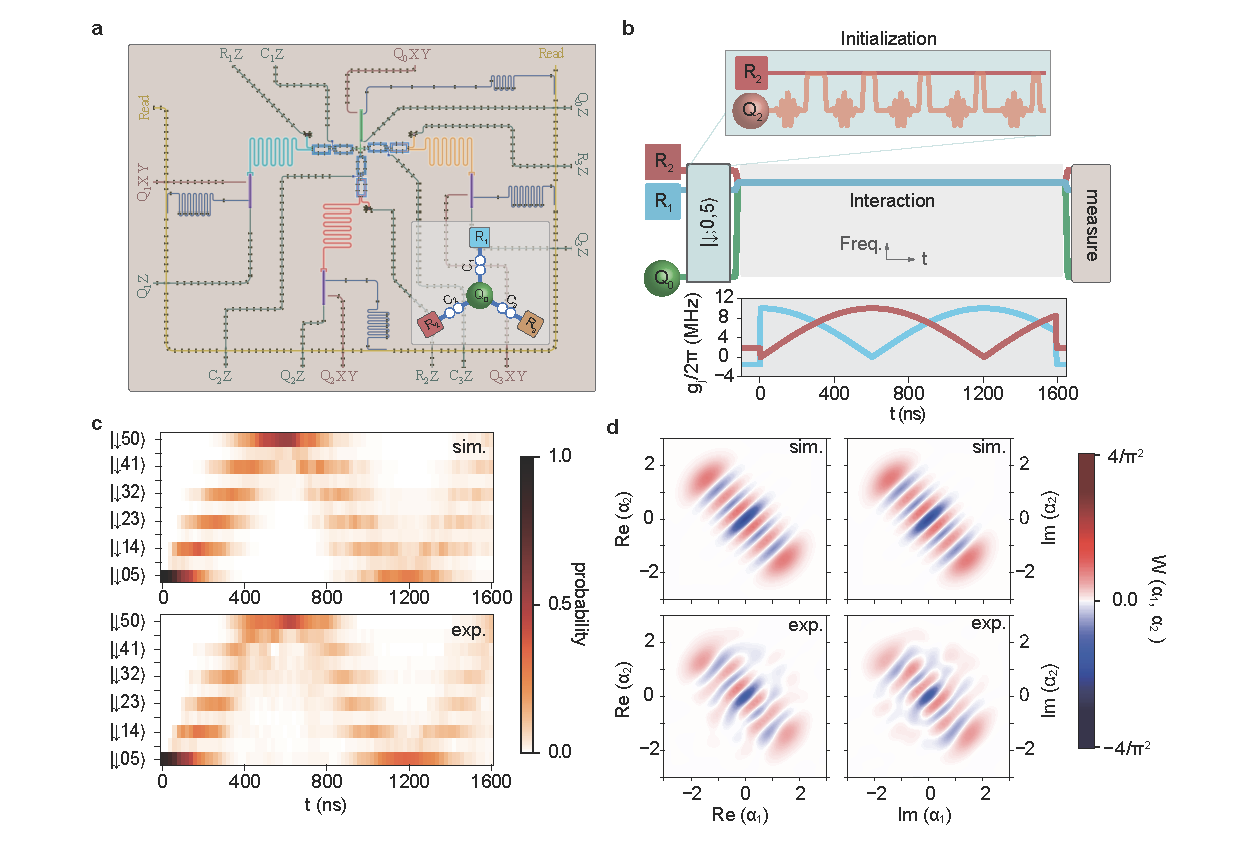}
		\caption{\label {Fig2}{\bf Adiabatic transport of the topological zero-energy states in the Fock-state Su-Schrieffer-Heeger model.} (a) False-color circuit image of the device fabricated and used in this experiment, where a central gmon qubit (green) is coupled to three resonators (R$_1$: cyan, R$_2$: red, R$_3$: yellow) via tunable couplers (blue). Each resonator R$_j$ is coupled to an ancilla qubit (purple). Inset shows the symbolized configuration of three resonators coupled to a qubit. (b) Experimental pulse sequences for the adiabatic transport. 
		We prepare the initial Fock state of R$_2$ by repeatedly exciting its ancilla qubit Q$_2$ with a $\pi$-pulse and tuning it in resonance with R$_2$ to swap the photons (upper panel). After the initialization we tune R$_1$, R$_2$ and Q$_0$ in resonance (middle panel) and modulate C$_1$, C$_2$ to tune the coupling strengths $g_1$ (cyan) and $g_2$ (red) (lower panel). Finally, we measure the joint-population of R$_1$, R$_2$ and Q$_0$. (c) The observed evolution of the zero-energy state wave packet in the numerical simulation (upper panel) and experiment (lower panel). Obviously $\left|\psi_0\right\rangle$ only occupies the $\left|\downarrow\right\rangle$ sublattice. In numerical simulation we use the parameters of the resonators and qubit listed in the Supplementary Table S1. All data {except quantum state tomography} in this paper are averaged over five runs of experiments. (d) The two-mode Wigner function of resonator state at $t = 300$ ns in the plane-cut along axes Re$(\alpha_1)$-Re$(\alpha_2)$ and Im$(\alpha_1)$-Im$(\alpha_2)$, whose fidelity is 0.735 (see Supplementary Information Fig.~S5 for pulse sequence of tomography and more data at other times).
		} 
	\end{figure*}

Leveraging the advantageous integrability and tunability of the circuit QED platform~\cite{Arute2019,Wu2021,Wang2019,Liu2020}, we design and fabricate a superconducting circuit device to build and engineer the FSLs.
The device contains a central gmon qubit~\cite{2016Chiral} (Q$_0$) and three resonators (R$_j$ with $j$ running from 1 to 3), all with tunable frequencies.~Each resonator R$_j$ 
is coupled to Q$_0$ through an inductive coupler (C$_j$), as illustrated in Fig.~\ref{Fig2} (a).
The coupling strengths, $g_j/2\pi$, can be continuously tuned from $-15$ MHz to 10 MHz by changing the magnetic flux in C$_j$.
In addition, each resonator R$_j$ is capacitively coupled to an ancilla qubit Q$_j$, which is used to prepare and readout the quantum state of the resonator, with a coupling strength of around $11$ MHz. 
Other characteristics of the resonators and qubits can be found in the Supplementary Information.


The Hamiltonian of the coupled system of R$_j$'s and Q$_0$ can be described by a multimode JC model \cite{Jaynes1963} in the rotating-wave approximation,
\begin{eqnarray}
\begin{aligned}
H = \frac{\hbar\omega_0}{2}\sigma_z+\sum_{j=1}^d\hbar\omega_ja_j^\dagger a_j+ \sum_{j=1}^d \hbar g_j(\sigma^+a_j +a_j^\dag  \sigma^-),
\end{aligned}
\end{eqnarray}
where $a_j$ is the annihilation operator of 
R$_j$ with the transition frequency $\omega_j$, $\sigma^+\equiv \left|\uparrow\right\rangle\left\langle\downarrow\right|$ and $\sigma^-\equiv \left|\downarrow\right\rangle\left\langle\uparrow\right|$ are the raising and lowering operators of Q$_0$ with the transition frequency $\omega_0$, and $d$ is the number of resonator modes. The Hamiltonian conserves the total excitation number $N=\sum_j n_j+(\sigma_z+1)/2$, where $n_j$ is the photon number of R$_j$ and $\sigma_z=\left|\uparrow\right\rangle\left\langle \uparrow\right|-\left|\downarrow\right\rangle\left\langle \downarrow\right|$. \\

\noindent\textbf{Topological transport}

\noindent When the number of modes $d=2$, in the subspace of $N$ excitations, there are $2N+1$ states $\left|s; n_1, n_2\right\rangle$ coupled in a bipartite tight-binding lattices with the spin states $s=\uparrow,\downarrow$ labelling the two sublattices, as shown in Fig.~\ref{Fig1} (a). When Q$_0$ is resonant with both resonators, all these $2N+1$ states have the same energy, which is set as the zero energy. Since the coupling strengths $t_j=g_j\sqrt{n_j}$ ($j=1,2$) depend on the photon numbers, we have $t_1>t_2$ and $t_1<t_2$ on the left- and right-hand sides of the FSL, resulting in two different topological phases of the SSH model \cite{Su1979}. 
	A topological zero-energy state locates around the lattice sites satisfying $t_1=t_2$, which is the topological edge of the SSH model. 
	We write $g_j=g_0\lambda_j$ where $g_0$ is a fixed non-zero coupling strength and $\lambda_j$'s are the tunable parameters satisfying $\lambda_1^2+\lambda_2^2=1$. The topological zero-energy state can be written as a two-mode binomial state \cite{Fu1997, Cai2020},
	\begin{eqnarray}
	\begin{aligned}
	|\psi_0\rangle = \sum_{n=0}^{N}\sqrt{\frac{N!}{n!(N-n)!}}\lambda_2^{n}(-\lambda_1)^{N-n}\left|\downarrow; n,N-n\right\rangle,
	\end{aligned}
	\label{psi}
	\end{eqnarray}
	which only occupies the $\left|\downarrow\right\rangle$ sublattice. By slowly tuning $\lambda_2$ from 0 to 1, we can adiabatically transport the zero-energy state from the right end of the lattice to the left end, or vice versa.

	
	\begin{figure}[t] 
		\includegraphics[width=3.2in]{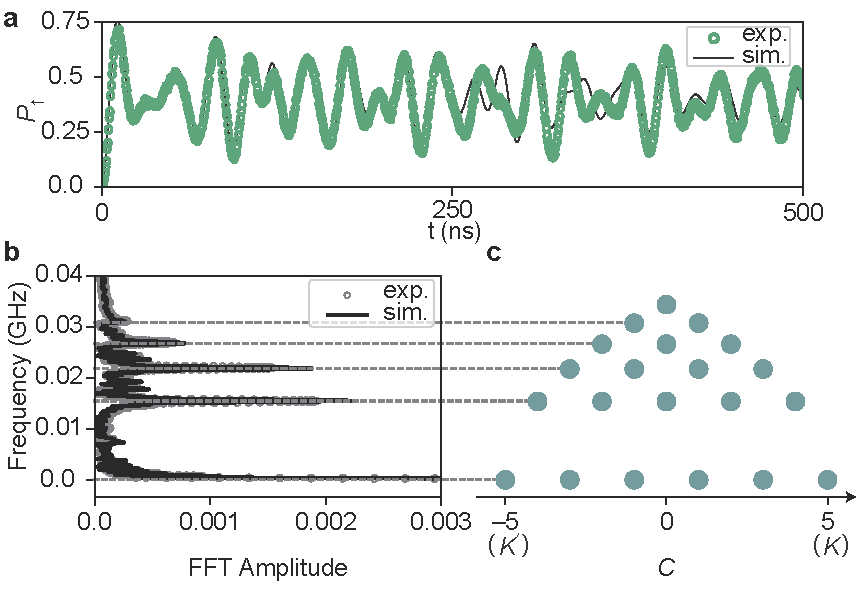}
		\caption{\label{Fig3}{\bf The pseudo-Landau levels in the two-dimensional Fock-state lattice with $N=5$.} (a) The evolution of the excited state population of the qubit. (b) Fast Fourier transform (FFT) of the Rabi oscillation. The vertical axis is the frequency component divided by two. The solid line is the numerical simulation and the circles are the experimental data. (c) The eigenstates in the pseudo-Landau levels of the 2D FSL, with eigenenergies corresponding to the FFT peaks. Each point labels an eigenstate characterized by the chirality $C$. The degeneracy of the $n$th Landau level is $N-n+1$.}
	\end{figure}
    In the experiment we select R$_1$ and R$_2$ to construct the SSH FSL,  with R$_3$ being far detuned and effectively decoupled from the system.
	The experimental pulse sequences are sketched in Fig.~\ref{Fig2} (b).
	We first prepare the initial state  $\left|\downarrow;0,5\right\rangle$, which is the topological zero-energy state of the SSH FSL with $N=5$ and $\lambda_1=1$, by pumping five photons successively into R$_2$ via Q$_2$ (upper pannel of Fig.~\ref{Fig2} (b)).
	Then we tune R$_1$, R$_2$ and Q$_0$ in resonance at the frequency $\omega_{\text{int}}/2\pi \approx 4.81$ GHz and sinusoidally modulate the coupling strengths where $g_0/2\pi \approx 9$ MHz, {$\lambda_1=|\cos(2\pi \nu t)|$ and $\lambda_2=|\sin(2\pi \nu t)|$} with $\nu= 416$ kHz $\ll g_0$ to satisfy the adiabatic condition, as shown in the lower panel of Fig.~\ref{Fig2} (b). 
	Finally the wave packet of the zero-energy state in the FSL is measured (see Methods), with the data shown in Fig.~\ref{Fig2} (c). The adiabatic transport of the topological edge state is witnessed by the oscillation of the photons between R$_1$ and R$_2$ following Eq.~(\ref{psi}) with time-dependent $\lambda_1$ and $\lambda_2$.
	The zero-energy state is topologically protected by the energy gap $g_0$ from other eigenstates of the FSL and maintains coherence during the transport \cite{Yuan2021a}. To show this, we further measure the density matrix of the two resonators by quantum state tomography. As shown in Fig.~\ref{Fig2} (d), the two-mode binomial state remains a Fock state in the combinational dark mode of the two resonators, $\lambda_2 a_1-\lambda_1 a_2$, and the quantumness of the states is evident from the negative values of the Wigner functions. \\
	

	\begin{figure*}[!htp]
		\includegraphics[width=7in]{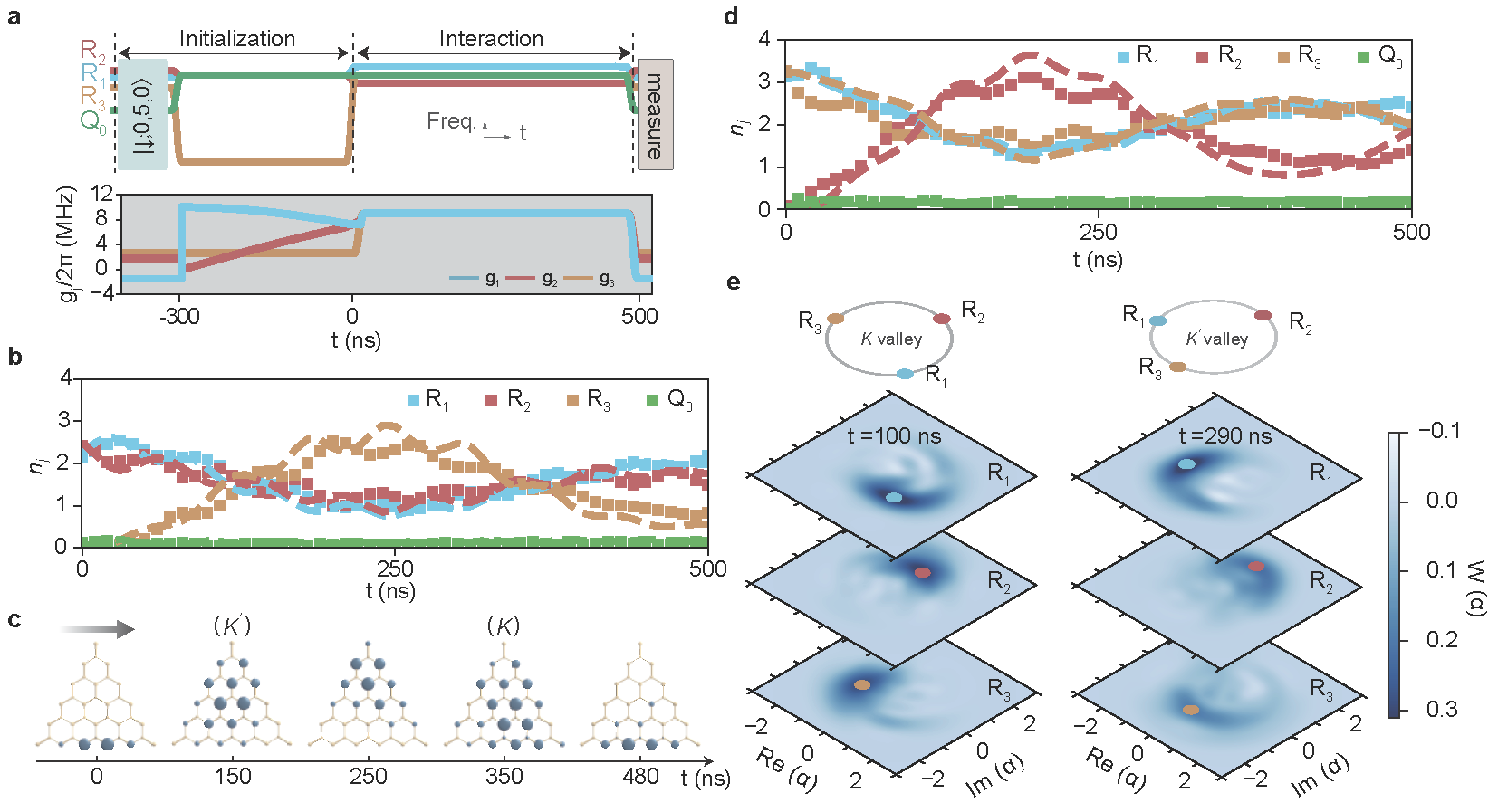}
		\caption{\label{Fig4} {\bf The valley Hall effect in the two-dimensional Fock-state lattice.} (a) The pulse sequences for controlling the frequencies (upper panel) and coupling strengths (lower panel). We first prepare the initial state $|\psi_0\rangle$ through the topological transport in the SSH FSL. 
		Then we tune R$_3$ and Q$_0$ in resonance at $\omega_\text{int}/2\pi \
		\approx 4.81$ GHz while we detune R$_1$ and R$_2$ to introduce the linear potential. Meanwhile we set the coupling strengths $g_j/2\pi \approx 9$ MHz for $j=1,2,3$, and finally we measure the joint populations at different times during the evolution. (b) The valley Hall evolution of the average photon numbers in the three resonators for an initial two-mode binomial state $|\psi_0\rangle$ for $N=5$. The squares are experimental data and the dashed lines are numerical simulations where we set the detuning $\delta/2\pi=1.8$ MHz. (c) The distribution of the valley Hall wave functions in the FSL for $N=5$ at $t=0$ ns, $150$ ns, $250$ ns, $350$ ns, $480$ ns. The left, right and top vertices correspond to the states with all photons in $R_1$, $R_2$ and $R_3$. The radius of the circle on each site is proportional to its population. The trajectory of the wave function is perpendicular to the direction of the effective force (black arrow). (d) The evolution of the average photon numbers in the three resonators for the coherent initial state $|\psi_c\rangle=\left|\downarrow; \alpha, 0, -\alpha\right\rangle$ with $\alpha \approx \sqrt{3.2}$. We detune $R_1$ and $R_3$ to introduce a linear potential $V=\hbar\delta(a_1^\dagger a_1-a_3^\dagger a_3)$. In the numerical simulation (dashed lines) we set $\delta / 2\pi = 2.35$ MHz. (e) The measured Wigner functions of the three resonator states at time $t=100$ and $290$ ns (see Supplementary Information Fig.~S3 for the numerical simulation). The phases of the largest amplitudes of the Wigner functions are labeled on the unit circles, which show the chirality of the corresponding states in the two valleys.} 
	\end{figure*}

\noindent\textbf{Valley Hall effect}

\noindent When $d=3$, the Fock states in the subspace with $N$ excitations form a two-dimensional honeycomb lattice of $(N+1)^2$ sites, as shown in Fig.~\ref{Fig1} (b). The site-dependent coupling strengths introduce a strain, which has the effect of a magnetic field and results in $\sqrt{n}$-scaling pseudo-Landau levels~\cite{Cai2020} when Q$_0$ is resonant with all three resonators. We observe the Landau levels by analyzing the spectra of the lattice dynamics \cite{Roushan2017}. In the experiment, we prepare the initial state $\left|\downarrow;0,5,0\right\rangle$ and resonantly couple R$_1$, R$_2$ and R$_3$ to Q$_0$ with coupling strengths $g_j/2\pi \approx 9$ MHz. We measure the evolution of the probability of finding Q$_0$ in the $\left|\uparrow\right\rangle$ state and then perform fast Fourier transform. We obtain peaks approximately located at $\sqrt{n}\Omega_0$ and {$\Omega_0 = \sqrt{3} g_j$}, as shown in Fig. \ref{Fig3} (c).  The degenerate states in the same Landau level are distinguished by their chiralities, 
\begin{eqnarray}
\begin{aligned}
	C=b^\dagger_+b_+-b^\dagger_-b_-,
\label{C}
\end{aligned}
\end{eqnarray}
where $b_\pm=\sum_{j=1}^3 a_j\exp{(\mp i 2j\pi/3)}/\sqrt{3}$ are the annihilation operators of the two chiral dark modes that are decoupled from the qubit. 
The chirality $C$ plays the role of the lattice momentum in conventional lattices and $C=N$ and $C=-N$ correspond to the two corners of the Brillouin zone, denoted as $K$ and $K^\prime$ valleys, respectively~\cite{Cai2020,RevModPhys.81.109}.

A Lifshitz topological edge \cite{Goerbig2011} locates on the incircle of the FSLs (see the dashed line in Fig.~\ref{Fig1} (b)). The expectation positions of the states in the zeroth Landau level are confined within the incircle by a band gap \cite{Cai2020}. 
The strain-induced magnetic field has opposite signs at the $K$ and $K^\prime$ valleys (see Supplementary Information).
By introducing a linear potential to mimic the effect of an electric field to electrons, we can observe the VHE (see Fig.~\ref{Fig1} (b)), i.e., the Hall response has opposite signs at the two valleys. To experimentally demonstrate this effect, we first prepare an initial state $|\psi_0\rangle$ with $\lambda_1=\lambda_2=1/\sqrt{2}$ in Eq.~(\ref{psi}) following the procedure in Fig.~\ref{Fig2} (b). Such an initial state on the Lifshitz topological edge is a Gaussian wave function in the zeroth Landau level \cite{Cai2020}. Then we bring R$_3$ in resonance with Q$_0$ and set {$g_j/2\pi \approx 9$ MHz} for $j=1,2,3$. The linear potential with horizontal gradient is introduced by slightly shifting the frequencies of R$_1$ and R$_2$, 
\begin{eqnarray}
\begin{aligned}
V= \hbar\delta (a_1^\dag a_1-a_2^\dag a_2),
\label{v}
\end{aligned}
\end{eqnarray}
where the detuning {$\delta/2\pi\approx 1.8$ MHz}.
We then measure populations on each lattice site in the FSL and obtain the average photon numbers in the three resonators, as shown in Fig.~\ref{Fig4} (b). The linear potential drives photons from $R_1$ and $R_2$ to $R_3$ while the qubit stays in the ground state.
To visualize the evolution of the wave function, we draw the population distributions in the FSL at five different times (see Fig.~\ref{Fig4} (c)).
The wave function first moves upward perpendicular to the force direction (black arrow) until being reflected by the Lifshitz topological edge near the top vertex, and then moves downward back to the initial state (up to a phase factor). In particular, when the wave function is at the center of the lattice but in different valleys, e.g., at $t=150$ and $350$ ns, it moves in opposite directions, which is a signature of the VHE \cite{Cai2020}. It is noteworthy that the qubit remains in the ground state during the evolution, which reflects a fundamental difference between classical and quantum optics (see Extended Data Fig.~\ref{extended_data_fig1}).

Surprisingly, the VHE can also be observed with initial classical states such as 
$|\psi_c\rangle=\left|\downarrow; \alpha, 0, -\alpha\right\rangle$, i.e., $R_1$ and $R_3$ are in the coherent states $|\alpha\rangle$ and $|-\alpha\rangle$ and $R_2$ is in the vacuum state. Such a state can be expanded as a superposition of two-mode binomial states similar to the initial state in Fig.~\ref{Fig4} (b) with different total excitation numbers $N$ \cite{Cai2020}. Since the dynamics different subspaces is synchronized, the fields in the three resonators remain as a direct product of coherent states and the evolution of the average photon numbers in the three resonators follow similar curves as those for the initial binomial state (see Fig.~\ref{Fig4} (d) and Supplementary Information). 

We can further measure the chirality of the states at the two valleys. Since the states of the three resonators are separable, we perform simultaneous quantum state tomography and obtain their Wigner functions (see Fig.~\ref{Fig4} (e)). As expected, the phases are distributed in a counter-clockwise ($C>0$) and clockwise ($C<0$) manner at $t=100$ and $290$ ns when the wave function moves to the $K$ and $K^\prime$ points, respectively. 
Therefore, the VHE in FSLs can be used to coherently transport the wave function between two valleys and control the chirality of the quantum states of multiple resonators.\\

\begin{figure}[t]
	\includegraphics[width=3.5in]{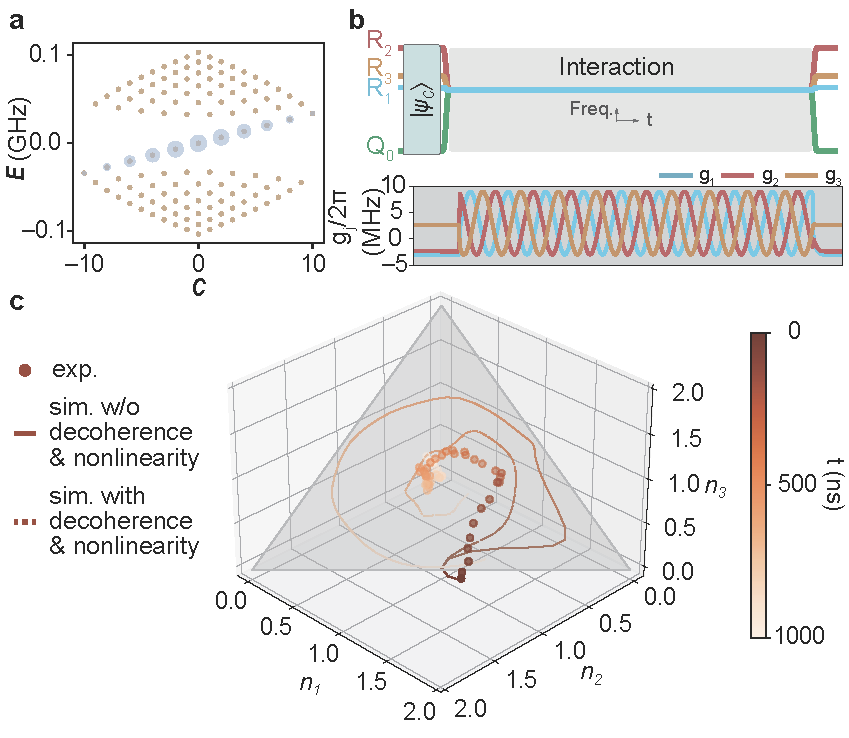}
	\caption{\label{Fig5} {\bf Chiral edge currents of the Fock-state Haldane model.} (a) The energy bands of the Hamiltonian $H_H$ in Eq.~(\ref{Haldane}) with total excitation number $N=10$, {$g_0/2\pi = 3.0$ MHz and $\kappa/2\pi=-0.54$ MHz}. Each yellow circle shows an eigenstate. There are chiral edge states connecting the two bands. The radii of the shaded circles on the chiral edge states are proportional to the population of a binomial state (with $\lambda_1=\lambda_2$) in the corresponding eigenstates. The initial state in our experiment has the same distribution in each subspace. (b) The control sequence in realizing the Haldane Hamiltonian. We prepare an initial state {$\left|\downarrow; \alpha, -\alpha, 0\right\rangle$} with $\alpha \approx 1$, {and tune three resonators and the gmon qubit on resonance at $\omega_{\text{int}}/2\pi \approx 4.73$ GHz}, followed by a Floquet modulation of  the coupling strengths $g_j(t)$ with static amplitude $g_0/2\pi \approx 3.0$ MHz, dynamic amplitude $g_d/2\pi \approx 3.0$ MHz and modulation frequency $\nu_d/2\pi=50$ MHz, such that the effective Haldane coupling strength $\kappa/2\pi=-0.54$ MHz. (c) Chiral edge currents shown by the average photon numbers in the three resonators. The total average photon number in the initial state is 2. The grey triangle shows the boundary of the FSL with $N=2$, which is the most occupied subspace at the initial time. The circles show the experimental data while the depths of the colors indicate the evolution time. The solid and dashed lines are the numerical simulations without and with resonator dephasing rate $\gamma_j/2\pi=0.25$ MHz and nonlinearity (see values in Supplementary Information). 
	To take into account the control imperfections, we slightly detune R$_1$ (R$_2$) by $0.5$ ($-0.8$) MHz in obtaining the dashed line for a better fitting.
	}
\end{figure}

\noindent\textbf{Haldane model}

\noindent By introducing a Floquet modulation of the coupling strength, $g_j(t)=g_0+2g_d\sin [\nu_d t+2(j-1)\pi/3]$, we synthesize a most important model in topological physics, the Haldane model, which has only been realized in cold atoms \cite{Jotzu2014} and very recently in moir\'e lattices \cite{Zhao2022}. The effective Hamiltonian in the second order perturbation is (see Supplementary Information),
	\begin{eqnarray}
	\begin{aligned}
	H_H= \hbar g_0 \sum_{j=1}^3(a_j^\dag \sigma_-+h.c.)+\hbar\kappa \sigma_z C,
	\label{Haldane}
	\end{aligned}
	\end{eqnarray}
where $\kappa=-3g_d^2/\nu_d$. The second term in Eq.~(\ref{Haldane}) introduces the complex next-nearest-neighbor hoppings in the FSL \cite{Wang2016, Cai2020} and transforms flat Landau levels to a two-band structure with gapless chiral edge states, which originate from the zeroth Landau level (see Fig.~\ref{Fig5} (a)). In the experiment we directly excite $R_1$ and $R_2$ to obtain an initial state $\left|\downarrow;\alpha,-\alpha,0\right\rangle$ (see Fig.~\ref{Fig5} (a) for its distribution in the subspace $N=10$, similar in other subspaces). Then we periodically modulate the coupling strengths $g_j(t)$ to realize the Haldane model (see the control sequence in Fig.~\ref{Fig5} (b)). The average photon numbers are subsequently measured as a function of time \cite{Saugmann2022AFS}, which shows the chiral motion, i.e., the wave function rotates in a counter-clockwise manner in the FSL (see Fig.~\ref{Fig5} (c)). Ideally the wave function shall be on the incircle, i.e., the Lifshitz topological edge. In the experiment the chiral rotating wave function moves toward the center of the FSL due to the decoherence and nonlinearity of the resonators.

In summary, we have demonstrated the coherent control of topological zero-energy states in 1D and 2D FSLs. These states only occupy the sublattice where the qubit is in the $\left|\downarrow\right\rangle$ state and they are protected from other eigenstates by an energy gap of the vacuum Rabi frequency. Perturbations with energy smaller than this gap, such as slow modulation of coupling strengths and small detunings between the resonators, can be used to coherently control the zero-energy state to realize topological transport, VHE and Haldane chiral edge currents. The techniques we have developed in this study can also be applied to control other eigenstates in the FSL, such as the excited states in higher Landau levels, although these states suffer from the decoherence of the qubit. With more resonators, we can synthesize higher dimensitional lattices that are not available in real space. Our study paves the way of investigating topological phases in FSLs and developing novel control methods for quantum state engineering of bosonic modes.

\vspace{.5cm}
\noindent\textbf{\large{}Methods}{\large\par}

\noindent\textbf{Device}

\noindent The device used in this experiment is composed of a top chip and a base chip that are assembled using the flip-chip bonding technique. The top chip integrates a total of 10 active circuit elements, i.e., 3 resonators, 4 qubits, and 3 inductive couplers, all of which are dynamically tunable with individual flux lines for Z controls. In addition, all qubits have their own microwave lines for XY controls. The base chip is connected with the top chip via In bumps, which are used to bridge different ground plane patches on the top chip. Details on the device fabrication can be found in~\cite{PhysRevLett.128.190502}.

As shown in Fig.~\ref{Fig2} (also see details in Fig. S1), the superconduncting resonators (R$_1$, R$_2$, and R$_3$) are interconnected by the central qubit (Q$_0$), where the resonator-qubit coupling strengths are dynamically programmable with the corresponding inductive couplers (C$_1$, C$_2$, and C$_3$). Each superconducting resonator R$_i$ ($i=1,2,3$) is a quarter-wavelength coplannar waveguide (CPW) resonator capacitively coupled to a standard frequency-tunable transmon qubit Q$_i$ for resonator-state control~\cite{PhysRevLett.101.240401, PhysRevLett.103.200404, nature_459_546, LinPeng_2013}. 
The center trace of R$_i$ is connected to an inductance coil and then to a two-junction superconducting quantum interference device (SQUID) before shorting to ground. The inductance coil is coupled to a coupler loop C$_i$. The SQUID with its own flux line provides the frequency tunability for the resonator and also leads to a slight nonlinearity {$\eta \approx -0.5$ MHz} of the resonator. Q$_0$ is a gmon qubit with its junction inductance being connected to an inductor tail, which is coupled to three coupler loops. 


The coupler loop consists of a Josephson junction and an inductance due to its loop geometry. The external flux across the coupler can be adjusted to change the effective inductance in order to tune the coupling strengths~\cite{2016Chiral}. An excitation in Q$_0$ (R$_i$) generates a current in the loop of C$_i$ which then excites the neighboring R$_i$ (Q$_0$). The magnetic flux through the coupler loop determines the effective junction inductance, which controls the current flowing through the coupler loop and effectively tunes the Q$_0$-R$_i$ coupling strengths.


\noindent\textbf{Resonator state preparation}

\noindent Two types of state, i.e., coherent states and Fock states, are used in this experiment. Starting from the initial vacuum state $|0\rangle$, the coherent state $|\alpha\rangle$ of the resonator is generated by resonantly driving it with a {60 ns} classical microwave pulse with controllable amplitude and phase. To prepare the Fock state $|n\rangle$ with $n\geq 1$, we repeatedly excite the ancilla qubit and bring it into resonance with the resonator for a time $\sim{\pi}/{2\sqrt{n}\Omega}$, where $\Omega$ is the coupling strength, to feed photons successively into the resonator~\cite{PhysRevLett.101.240401, nature_459_546}.

\noindent\textbf{Resonator population measurement}

\noindent To measure the population of a resonator mode, we bring the ancilla qubit, which is initialized in ground state $\left|\downarrow\right\rangle$, into resonance with the mode for a specific time $\tau$ and measure the $\left|\uparrow\right\rangle$-state probability $P_{\uparrow}(\tau)$ of the qubit.
The population can be obtained by fitting $P_{\uparrow}(\tau)$ with the formula 
 \begin{eqnarray}\label{qrswap}
\begin{aligned}
P_{\uparrow}(\tau) = \frac{1}{2}\sum_{n=0}^mP_{n} (1-\cos(2\sqrt n\Omega\tau )),
\end{aligned}
\end{eqnarray}
where $m$ is the highest occupied energy level of the resonator, $P_{n}$ is the population of the Fock state $|n\rangle$, and $\Omega$ is the coupling strength.~To measure the joint population of multiple resonator modes, we simultaneously bring the ancilla qubits into resonance with their corresponding resonators for a specific time $\tau$ and measuring the joint probabilities $\{P_{\downarrow\downarrow...\downarrow}(\tau), P_{\uparrow\downarrow...\downarrow}(\tau), ..., P_{\uparrow\uparrow...\uparrow}(\tau)\}$. During the interaction, the resonator modes are decoupled from each other, and the joint probability is
 \begin{eqnarray}\label{extended}
 \begin{aligned}
P_{s_1s_2...s_d}(\tau) = &\frac{1}{2^d}\sum_{n_1n_2...n_d}P_{n_1n_2...n_d} \times
\\ &\Pi_{i=1}^{d}[1+f(s_i)\cos(2\sqrt n\Omega_i\tau)],
 \end{aligned}
 \end{eqnarray}
where  $d$ is the number of resonator modes, $s_i=\downarrow,\uparrow$ and $n_i=0,1,2,...,m$ represents the state of the $i^{\text{th}}$ ancilla qubit and resonator mode, $P_{n_1n_2...n_d}$ is the joint population of $|n_1n_2...n_d\rangle$, and $\Omega_i$ is the coupling strength of the $i^{\text{th}}$ qubit-resonator pair. 
$f(s)=1$ and $-1$ for $s=\downarrow$ and $\uparrow$. For $d=1$, Eq.~(\ref{extended}) reduces to Eq.~(\ref{qrswap}). We obtain the joint population via least-squares optimization.

\noindent\textbf{Resonator state tomography}

\noindent We characterize the density matrix of the resonator mode by applying resonator tomography, which is achieved by inserting a set of displacement operations $D(\beta)=e^{\beta a^\dagger-\beta^*a}$ to the resonator before measuring the population. 
Each displacement operation relocates the off-diagonal terms of the density matrix to the diagonal terms of the resulting new matrix, which can be measured experimentally.
By choosing a set of $\{\beta_k\}$ and measuring the corresponding population, the original density matrix can be reconstructed from a set of over-constrained equations via the least-squares optimization (see~\cite{LinPeng_2013} for details).
In this experiment, $\beta$ are sampled from three concentric circles with different radii.
Similarly, we can characterize the density matrix of multiple modes by applying on them a set of displacement operators selected from $\{\beta_k\}^{\otimes d}$ simultaneously, with $d$ being the number of modes, and measuring the joint population.

\vspace{.6cm}
\noindent\textbf{\large{}Data availability}
The data presented in the figures and that support the other findings of this study will be uploaded to some open repositories and will be publicly available upon the publication of the paper. 

\vspace{.5cm}
\noindent\textbf{Acknowledgement} We thank Ke Wang, Wenhui Ren, Jiachen Chen and Xu Zhang for technical support during the experiment. The device was fabricated at the Micro-Nano Fabrication Center of Zhejiang University. We acknowledge the support of the the National Key Research and Development Program of China (Grant No.~2019YFA0308100), National Natural Science Foundation of China (Grants No.~11934011, No.~12174342, No.~92065204 and No.~U20A2076), the Zhejiang Province Key Research and Development Program (Grant No. 2020C01019), and the Fundamental Research Funds for the Zhejiang Provincial Universities (Grant No. 2021XZZX003).

\vspace{.3cm}
\noindent\textbf{Author contributions}  
D.-W.W., C.S. and H. W. designed the experiment; J.D.~and H.D. designed the device and carried out the experiments supervised by C.S.~and H.W.; H.L.~fabricated the device supervised by H.W.; J.D., C.Z., Y.W.~and J.Y.~performed the numerical simulations supervised by C.S.~and D.-W.W.; J.D., C.S. and D.-W.W. wrote the manuscript with comments and inputs from other authors; All authors contributed to the analysis of data and the discussions of the results.

\vspace{.3cm}
\noindent\textbf{Competing interests}  All authors declare no competing interests.

\vspace{.3cm}
\noindent{* These authors contributed equally to this work.\\
$^\dagger$ chaosong@zju.edu.cn\\
$^\S$ hhwang@zju.edu.cn \\
$^\ddagger$ dwwang@zju.edu.cn}

\bibliography{FSL}

\renewcommand{\figurename}{Extended Data Fig.}
\setcounter{figure}{0}  
\begin{figure*}[!htbp]
	\includegraphics[width=5in]{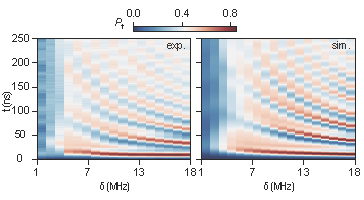}
	\caption{\label {extended_data_fig1} {\bf Transition from the classical to quantum regimes.} The experimental setup is similar to the one in Fig.~\ref{Fig4} for the valley Hall effect. In Fig.~\ref{Fig4} the atom stays in the ground state for a small $\delta$. Here we show the evolution of population in the qubit excited state $P_\uparrow$ for different values of $\delta$. For an initial coherent states $\left|\alpha, 0, -\alpha\right\rangle$ being coupled to the qubit with the linear potential $V=\hbar\delta(a_1^\dagger a_1-a_3^\dagger a_3)$, classical optics and quantum optics give distinct predictions for the evolution of the coupled system. In classical optics, the qubit is coupled to a superposition classical field with amplitude proportional to $|\alpha|\sin(\delta t)$, such that the qubit undergoes Rabi oscillation with time-dependent frequencies. This classical regime is shown by the fringes in $P_\uparrow(t)$ for large $\delta$. In this regime the detuning $\delta$ is large enough to induce transitions between Landau levels. By decreasing $\delta$ we can observe the the transition from classical to quantum regimes. In particular, when $\delta < 3g_1/N$, the total potential difference across the incircle is smaller than the band gap between the zeroth and first Landau levels, such that the transitions between Landau levels are inhibited. The dynamics of the zero-energy states is dominated by the valley Hall effect within the zeroth Landau level (which only occupies the $\left|\downarrow\right\rangle$-state sublattice, see Fig.~\ref{Fig4}) and the Rabi oscillation disappears. In this experiment the boundary is at about $\delta/2\pi=3$ MHz below which the (quantum) valley Hall response dominates and above which the (classical) Rabi oscillation dominates. Here we set $\alpha \approx 2$ in the experiment.}
\end{figure*}

\end{document}